# Variation on a theme by GHZM

P.K.Aravind
Physics Department
Worcester Polytechnic Institute
Worcester, MA 01609
(email: paravind@wpi.edu)

ABSTRACT

In the Greenberger-Horne-Zeilinger-Mermin (GHZM) proof of Bell's theorem, a source periodically emits an entangled state of three particles whose properties are analyzed by three distant observers and used to prove Bell's nonlocality theorem. This paper analyzes a somewhat different *gedanken* experiment involving only two observers that nevertheless makes indirect use of the GHZ states to prove Bell's theorem. The relationship of the GHZM proof to the present one is discussed, and it is pointed out that the latter provides an interesting new view of the connection between the "two theorems of John Bell".

> **Faust**: The Pentagram embarrasses you? Tell me then thou child of hell, if that repels thee, how cam'st thou in? How was such a spirit entrapped?
> **Mephistopheles**: The poodle observed nothing when he jumped in. The thing looks differently now; the devil cannot get out.
> *Goethe's Faust*

**I. The magic pentagrams …**

Alice and Bob, that dynamic duo, have decided to put on yet another magic show to puzzle and delight their fans. This show is the latest in a line of tricks they have performed with the help of David Mermin [1-3]. In all these shows, a central source periodically emits bursts of particles that cause several detectors at some distance from it to flash red and green lights in a manner that seems to defy any logical explanation. This latest trick is actually a variation of one of the earlier tricks in the series due to GHZ [4], as modified and presented by M [2], but it may not be obvious even to those who have witnessed (and fathomed) that trick. Accordingly, it seems worth presenting for whatever surprise it may occasion and the consequent educational value it may have. I proceed to the description of the trick without further ado.

A central source S has a button that, when pressed, causes it to emit particles that travel to the left and the right and impinge on two detectors A and B at some distance from it (see top part of Fig.1). Detector A is operated by Alice and detector B by Bob, and the two detectors are identical in all respects. Each detector has a display panel consisting of ten light bulbs arranged in the form of a pentagram (see bottom part of Fig.1), with four bulbs lying along each edge of the pentagram and each bulb lying at the intersection of two edges. The edges of the pentagram are labeled S1, S2, S3, S4 and S5, as shown in Fig.1. Each detector has a switch that can be set to one of five positions, also labeled Si ($i = 1,..,5$), each of which causes each of the bulbs lying along the corresponding edge of the pentagram to light up either red or green when the particles from the source enter that detector. With this explanation of the setup, we are ready to present the magic show. The show consists of a large number of repetitions of the following three steps (which we will refer to collectively as a "run"):

(i) the button is pressed on the source, causing it to release its particles towards the detectors;
(ii) after the button is pressed, but before the particles reach their detectors, Alice and Bob independently and randomly set their detectors to one of the allowed switch settings.
(iii) when the particles reach their detectors and activate them, Alice and Bob each record the colors of the lights that flash on the bulbs picked out by their respective switch settings.

Step (ii) must be performed with some care to ensure that the show is not later dismissed as a hoax. To avoid this charge it is necessary that, in any run, Alice's switch setting and the flashing of her detector lights both occur at spacelike separations from both the corresponding events at Bob's end. This condition guarantees that neither of the key events at either person's end (i.e. the choice of switch setting or the colors that flash on the detector panel) have any effect on the corresponding events at the other person's end. (The possibility of exchanging signals between the two ends, which some viewers may be tempted to invoke as an explanation for the magic about to be revealed, thereby gets ruled out.)

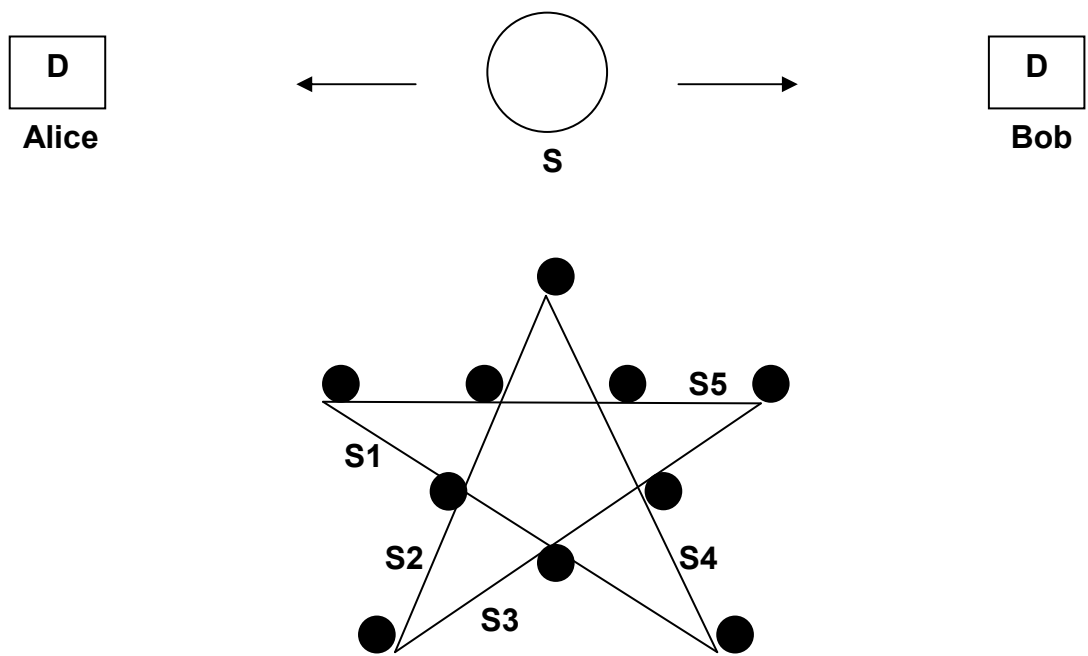

Fig.1. The top part of the figure shows a central source, S, containing a button that, when pressed, causes it to emit particles towards two detectors, labeled D, on either side of it. The detector on the left is operated by Alice and that on the right by Bob. The lower part of the figure shows the pentagrammic array of ten bulbs visible in the display panel of each detector. The bulbs fall into five lines of four bulbs each, with each line coinciding with an edge of the pentagram and bearing the label S1, S2, S3, S4 or S5, as shown. When a particular switch setting Si (i = 1,2,3,4 or 5) is chosen on a detector, the bulbs lying along that edge of the pentagram each light up either red or green when the particles from the source enter that detector.

After a large number of runs of the show, Alice and Bob get together to compare their records of switch settings and light flashings. They then find that in every one of their runs their detector outputs always conform to the following two rules:

I. <u>Parity Rule</u>: The number of bulbs that lights up red on a detector is always odd (and hence so is the number that lights up green).

II. <u>Correlation Rule</u> : Any bulbs common to both Alice's and Bob's switch settings always light up the same color(s) on both detectors.

Figure 2 shows some typical runs from a show. Note that the parity and correlation rules are obeyed in every run (the former by each of the detectors individually, and the latter by them both jointly).

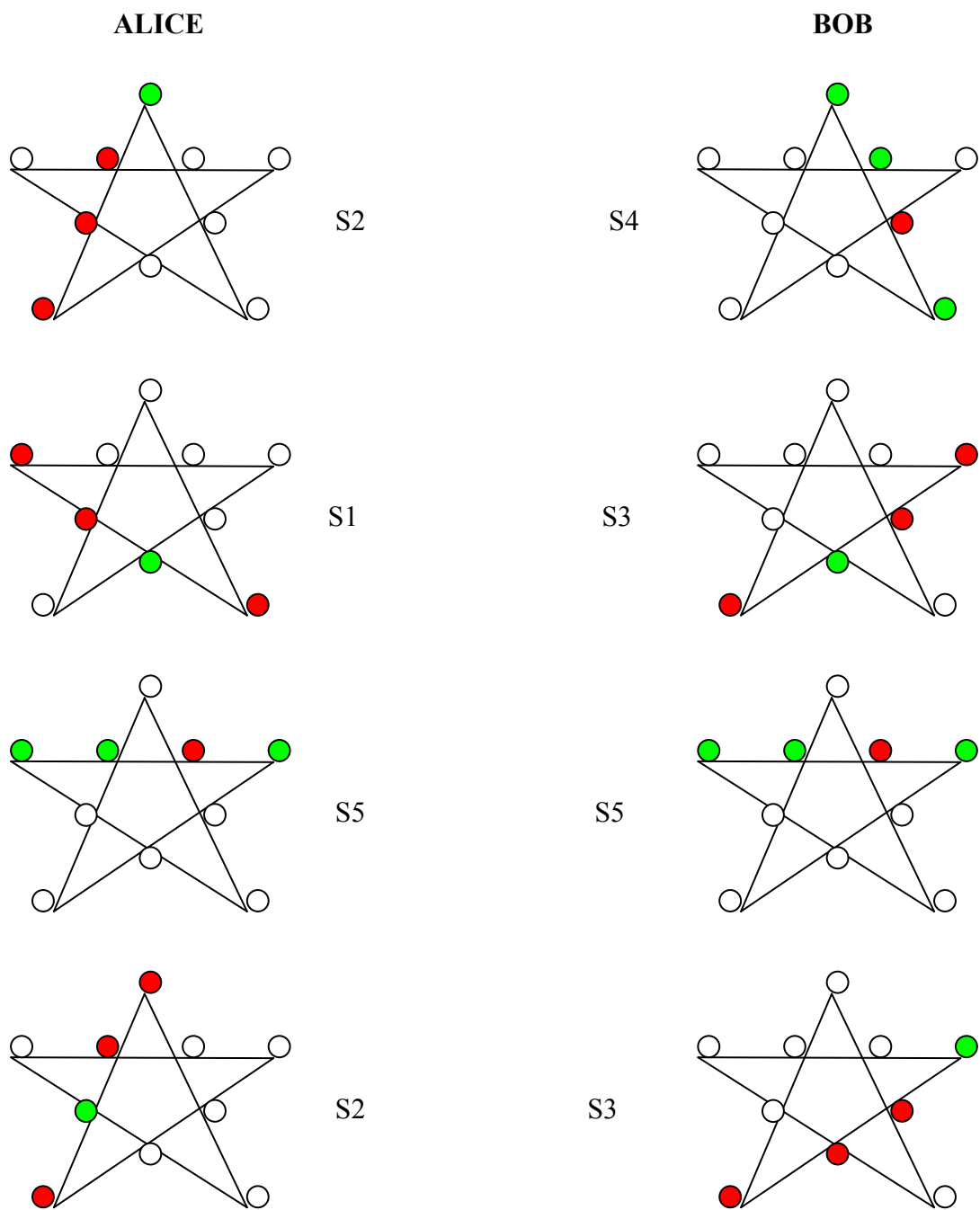

Fig.2. Alice's and Bob's switch settings and detector outputs shown from top to bottom for four different runs.

What is so magical about the results of this show, as revealed in Fig.2? The answer is that it is impossible to give any intuitively plausible explanation of how the parity and correlation rules can both be invariably obeyed in every run. Let us see why.

The only way in which the correlation rule can always be satisfied, if one rules out the possibility of any exchange of information between the two ends, is if the particles going to the two detectors in any run always give them identical "instructions" on how to light up for each of their switch settings; further, the instructions must make each bulb light up the same color no matter through which of the two possible switch settings it is activated, for only then can the correlation rule be satisfied whether Alice and Bob choose the same switch setting or not. In other words, satisfaction of the correlation rule requires that each detector be given an "instruction set" in any run telling it what color to light up each of its ten bulbs, and also that the instruction sets given to the two detectors be the same. It remains only to see how such instruction sets can be constructed in conformity with the parity rule. But it is here that an obstacle looms. Let $n_i$ be the number of red bulbs (i.e. bulbs instructed to light up red) in the line Si (i=1,..,5) of a general instruction set, and let $N = n_1 + n_2 + n_3 + n_4 + n_5$. Then, on the one hand, $N$ must be odd (because the parity rule requires each $n_i$ to be odd) but, on the other hand, $N$ must also be even (because each red bulb occurs at the intersection of two lines and so is counted in two of the $n_i$'s). This contradiction shows that instruction sets consistent with both the parity and correlation rules are impossible, making the results obtained by Alice and Bob seem profoundly mysterious.

**II. … and how they work**

Pressing the button on the source causes it to emit three Bell states, with one member of each Bell state going to Alice and the other to Bob. More precisely, the source emits the state

$$|\Psi\rangle = \frac{1}{\sqrt{2}}(|0\rangle_1|0\rangle_2 + |1\rangle_1|1\rangle_2) \otimes \frac{1}{\sqrt{2}}(|0\rangle_3|0\rangle_4 + |1\rangle_3|1\rangle_4) \otimes \frac{1}{\sqrt{2}}(|0\rangle_5|0\rangle_6 + |1\rangle_5|1\rangle_6), \quad (1)$$

with qubits 1,3 and 5 going to Alice and qubits 2,4 and 6 to Bob ($|0\rangle$ and $|1\rangle$ are the standard computational basis states of a qubit, i.e. they are the eigenstates of the Pauli operator $\sigma_z$ with eigenvalues +1 and -1, respectively.)

Figure 3 shows ten observables for a system of three qubits arranged at the vertices of a pentagram, with the four observables along each edge forming a mutually commuting set. Each observable has eigenvalues $\pm 1$ and the product of the four observables along any edge is $-I$, where $I$ is the identity operator. When the switch setting Si (i = 1,..,5) is chosen on a detector, the detector carries out a measurement of the four commuting observables along the corresponding edge on the three qubits entering it and displays the results on the bulbs associated with those observables according to the convention that an eigenvalue of +1 is displayed as a green light and one of -1 as a red light. The parity rule is then an immediate consequence of the fact that the product of the eigenvalues of the observables along any edge of the pentagram is -1.

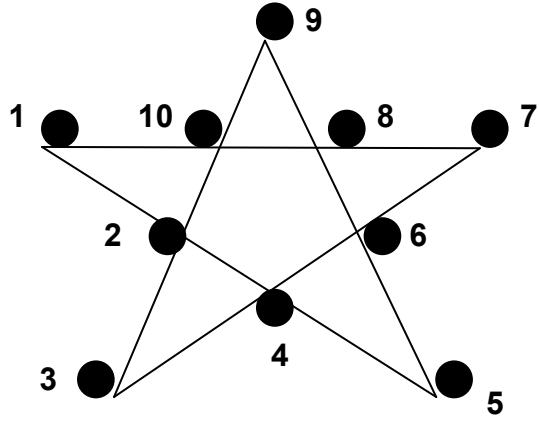

Fig.3. Ten observables for a system of three qubits, arranged at the vertices of a pentagram. The observables are $1 = -\sigma_z\sigma_z\sigma_z$, $2 = \sigma_z II$, $3 = I\sigma_x I$, $4 = II\sigma_z$, $5 = I\sigma_z I$, $6 = \sigma_x II$, $7 = -\sigma_x\sigma_x\sigma_z$, $8 = -\sigma_x\sigma_z\sigma_x$, $9 = II\sigma_x$ and $10 = -\sigma_z\sigma_x\sigma_x$, where $\sigma_x, \sigma_y, \sigma_z$ and $I$ are the Pauli and identity operators of a qubit. The first, second and third observables in each product refer to qubits 1,3 and 5 of state (1) for Alice and to qubits 2,4 and 6 of the same state for Bob. Alice and Bob get these qubits of state (1) in every run and carry out a measurement of a set of commuting observables (along one of the edges of the pentagram) on them. The results are displayed on the bulbs associated with the observables according to the convention that an eigenvalue of +1 (or -1) is shown as a green (or red) light.

The origin of the correlation rule can be understood as follows. Let $|\psi_i\rangle (i=1,..,8)$ be an arbitrary set of orthonormal states in the space of qubits 1,3 and 5 whose expansion coefficients in the standard basis $\{|000\rangle, |001\rangle, ..., |111\rangle\}$ are all real. Let $|\phi_i\rangle (i=1,..,8)$ be an identical set of orthonormal states (the "partner" states) in the space of qubits 2,4 and 6. It is easy to verify that the state (1) can be expressed in terms of these two sets as

$$|\Psi\rangle = \frac{1}{\sqrt{8}}\left[|\psi_1\rangle|\phi_1\rangle + |\psi_2\rangle|\phi_2\rangle + ... + |\psi_8\rangle|\phi_8\rangle\right]. \qquad (2)$$

Suppose now that Alice carries out a measurement on her qubits of one of the sets of commuting observables in Fig.3. Let $|\psi_i\rangle (i=1,..8)$ be the simultaneous eigenstates of Alice's set of measurement observables and let $|\phi_i\rangle$ be the corresponding partner states in the space of Bob's qubits. The source state can be expressed in terms of these states as shown in (2). When Alice carries out her measurement on her qubits, the state (2) collapses (with equal likelihood) into one of the product states $|\psi_i\rangle|\phi_i\rangle$ of which it is made up [5], and Alice observes the combination of eigenvalues associated with the state $|\psi_i\rangle$. An expansion of the state $|\phi_i\rangle$ in terms of the complete, orthonormal set of eigenstates associated with Bob's measurement observables shows that any of Bob's observables in common with Alice's is/are forced to return the same eigenvalue(s) for him as for her – which is just the correlation rule. The same conclusion also follows if one repeats the analysis with Bob's measurement preceding Alice's.

**III. GHZM vs the present trick**

It is instructive to compare the present trick with the GHZM one, as there may be something to be learned in the process. One conspicuous difference between the two tricks is that GHZM makes use of three observers while the present scheme requires only two. However GHZM makes up for the extra observer by having the observers carry out measurements on only a single qubit, whereas the present scheme requires the observers to carry out a somewhat complicated measurement of three commuting 3-qubit observables if they choose the switch setting S5. The challenge of carrying out the S5 measurements makes the present scheme more difficult to realize in the laboratory than the GHZM scheme. However the expertise required to carry out such measurements is now available in a few laboratories, such as Prof. Zeilinger's, and so an experimental realization of this scheme does not appear to be out of the question.

As far as proving Bell's theorem is concerned, GHZM seems to have a bit of an edge over the present scheme. This is because its measurements fall neatly into two classes: a set of preparatory measurements that establish that the spin components of the individual qubits are "elements of reality", followed by a final climactic measurement in which the product of the three spin components is found to have the opposite value from that predicted by local realism. The present scheme does not offer a dramatic climax of this sort but claims its victory after a long series of runs in which the parity and correlation rules are consistently upheld. An outright contradiction is always more striking than a steadily accumulating mountain of evidence, so GHZM will be found more convincing by many. However it should be pointed out that the present scheme is more compelling than a purely statistical test because its repudiation of local realism can be seen in every run, rather than in a statistical pattern abstracted from a large number of runs.

The role of the GHZ state in the two schemes is also worth contrasting. In the GHZM scheme, the GHZ state is emitted by the source and so plays a starring role throughout the action. By contrast, the GHZ state is not emitted by the source in the present scheme but it puts in a cameo appearance when one, or both, of the detectors fire in response to the switch setting S5. As a further twist, not just one GHZ state, but all members of a complete orthonormal set of eight, can put in this cameo appearance, as evidenced by the occurrence of all eight possible color combinations of the bulbs in the line S5 over the course of the experiment. Thus, although the GHZ state has been relegated to the background in the present scheme, it makes up for this by putting in a considerably more colorful (and unanticipated) appearance.

The most striking contrast between the GHZM and present schemes is in the approach they take to establishing the "two theorems of John Bell" [6]. One of these theorems, due to Bell [7] and to Kochen and Specker [8], rules out the existence of noncontextualist hidden variable theories while the other more famous one, due to Bell [9], rules out the existence of local hidden variable theories. The GHZM approach to proving these theorems was laid out by Mermin [6], who used the pentagram of Fig.3 to prove the Bell-Kochen-Specker (BKS) theorem and then passed on to a source emitting GHZ states and a trio of observers to prove Bell's theorem. This was a very impressive performance, but it required a short pause between the first act (BKS) and the second act (Bell), while the scenery was changed a little bit and two additional actors were brought in. The present scheme, by contrast, uses a "double play" to prove both theorems within the same general setting: Alice and Bob first each prove the BKS theorem for themselves by noting that their detector outputs always obey the parity rule; then they get together and compare their observations and conclude, from the fact that the correlation rule is always obeyed, that they have also proved Bell's theorem. It is this practically seamless link between the proofs of the BKS and Bell theorems that is the most striking feature of the present scheme.

The fact that the BKS theorem can be used as a "catalyst" in proving Bell's theorem, when used in conjunction with entanglement in the right manner, has been noted earlier in special cases by Heywood and Redhead [10], Zimba and Penrose [11] and perhaps others. The author tried to make this point in all generality in ref.[12].

The work of GHZM represents a truly significant extension of the deep and far reaching ideas of John Bell. The fact that it admits of the variation presented here is indicative of its power to shed light on the multifaceted, and still elusive, phenomenon of entanglement.

**Endnote.** This paper is dedicated informally to Daniel Greenberger, Michael Horne, Anton Zeilinger and David Mermin. I thank Martin Gardner for directing my attention to the lines from Goethe's Faust quoted at the beginning of this paper (the lines are from the second section of Faust, entitled "Study", and they are taken from A.Hayward's prose translation of 1882).

REFERENCES

1. N.D.Mermin, "Bringing home the atomic world: Quantum mysteries for anybody", Am. J. Phys. **49**, 940-3 (1981). An expanded version of this paper can be found as Ch.12 in N.D.Mermin, *Boojums all the way through* (Cambridge U.P., Cambridge, 1990). See also N.D.Mermin, "Is the moon there when nobody looks? Reality and the quantum theory", Phys. Today **38** (4), 38-47 (1985).

2. N.D.Mermin, "Quantum mysteries revisited", Am. J. Phys. **58**, 731-4 (1990). This is a nontechnical account of the GHZ proof of Bell's theorem in ref.4.

3. N.D.Mermin, "Quantum mysteries refined", Am. J. Phys. **62**, 880-7 (1994). This is a nontechnical account of Hardy's proof of Bell's theorem in L.Hardy, "Non-locality for two particles without inequalities for almost all entangled states", Phys. Rev. Lett. **71**, 1665-8 (1993).

4. D.M.Greenberger, M.Horne and A.Zeilinger, "Going beyond Bell's theorem", in *Bell's Theorem, Quantum Theory, and Conceptions of the Universe*, edited by M.Kafatos (Kluwer Academic, Dordrecht, 1989) pp.69-72; D.M.Greenberger, M.Horne, A.Shimony and A.Zeilinger, "Bell's theorem without inequalities", Am. J. Phys. **58**, 1131-43 (1990).

5. This is the content of the generalized Born rule. See, for example, N.D.Mermin, *Lecture Notes on Quantum Computation*, Chapter 1, available at
http://people.ccmr.cornell.edu/~mermin/qcomp/CS483.html

6. N.D.Mermin, "Hidden variables and the two theorems of John Bell", Rev. Mod. Phys. **65**, 803-15 (1993).

7. J.S.Bell, "On the problem of hidden variables in quantum mechanics", Rev.Mod.Phys. **38**, 447-52 (1966). Reprinted in J.S.Bell, *Speakable and Unspeakable in Quantum Mechanics* (Cambridge University Press, Cambridge, New York, 1987).

8. S.Kochen and E.P.Specker, "The problem of hidden variables in quantum mechanics", J. Math. Mech. **17**, 59-88 (1967).

9. J.S.Bell, "On the Einstein-Podolsky-Rosen paradox", Physics *1*, 195-200 (1964). Reprinted in the book quoted in Ref.6.

10. P.Heywood and M.L.G.Redhead, "Nonlocality and the Kochen-Specker paradox", Foundations of Physics **13,** 481-499 (1983).

11. J.Zimba and R.Penrose, "On Bell nonlocality without probabilities: more curious geometry", Studies in History and Philosophy of Science **24**, 697-720 (1993).

12. P.K.Aravind, "Impossible colorings and Bell's theorem", Phys. Lett. A262, 282-6 (1999). P.K.Aravind, "Bell's theorem without inequalities and only two distant observers", Found. Phys. Lett. **15**, 399-405 (2002).